\documentstyle[10pt,aaspp4,flushrt]{article}

\begin{document}

\title{Inertia of Heat in Advective Accretion Disks\\
       around Kerr Black Holes} 

\author{A.M. Beloborodov\altaffilmark{1,}\altaffilmark{2,}\altaffilmark{3}, 
M.A. Abramowicz\altaffilmark{3,}\altaffilmark{4}, and
I.D. Novikov\altaffilmark{5,}\altaffilmark{2,}\altaffilmark{1,}\altaffilmark{4}}
\altaffiltext{1}{Astro-Space Center of Lebedev Physical Institute, 
Profsoyuznaya 84/32, Moscow 117810, Russia} 
\altaffiltext{2}{Theoretical Astrophysics Center, Julian Maries Vej 30, 
DK-2100 Copenhagen, Denmark}
\altaffiltext{3}{Chalmers University of Technology, 412 96 Goteborg, 
Sweden}
\altaffiltext{4}{Nordita, Blegdamsvej 17, DK-2100 Copenhagen, Denmark}
\altaffiltext{5}{Copenhagen University Observatory, Julian Maries Vej 30, 
DK-2100, Copenhagen, Denmark}

\begin{abstract}

In the innermost region of the advective accretion disk orbiting
a black hole of high spin, the inertia of heat stored in the
accreting gas is comparable to that of the gas rest mass itself. 
Accounting for this effect, we derive additional terms
in the disk  structure equations, and show that the heat
inertia plays a significant role in the global energy conservation
and dynamics of accretion in the relativistic advective disks.

\end{abstract}


\section{Introduction}

In advective accretion disks (see Narayan 1996 for a review), 
the released gravitational energy is not
radiated away as in the standard model of thin disk (Shakura 1972; 
Shakura \& Sunayev 1973). Instead, it is stored in the form of internal heat, 
being eventually advected into the black hole.
Optically thin advective disks have been applied
to X-ray transients, low luminosity active galactic nuclei, and the Galactic
Center (Narayan 1996).
Accretion can also proceed in the advection dominated regime in the 
optically thick super-Eddington accretion disks (Abramowicz {\it et al.} 1988), 
and the observational significance of these disk models for quasars
has been recently discussed (e.g. Szuszkiewicz, Malkan, \& Abramowicz 1996). 

Most of the models of the advective disk have been constructed under
assumptions similar to those used in the standard model: 
that the flow is stationary and axially symmetric, and that one
dimensional approximation, with the ``integrated'' vertical structure,
is physically adequate. At the same time, the equations of such a {\it slim} 
disk, differ considerably from the standard ones. 
In particular, the slim radial Euler equation accounts for a deviation of gas 
motion from the Keplerian one, which becomes significant in the case of 
advection dominated disk. A Newtonian version of
the slim disk equations has been derived by Paczy{\'n}ski and
Bisnovatyi-Kogan (1981) long time ago and used by many authors.  
The relativistic version of these equations was derived
more recently by Lasota (1994; see also Abramowicz {\it et al.} 1996;
Abramowicz, Lanza, \& Percival 1997; Peitz \& Appl 1997).
Lasota's equations have been integrated numerically to obtain 
relativistic models for the hot disk of low luminosity
(Abramowicz {\it et al.} 1996; Jaroszy{\'n}ski \& Kurpiewski 1997; 
Peitz \& Appl 1997; Igumenshchev, Abramowicz, \& Novikov 1997).

Accreting black holes are expected to acquire a large spin and then the 
relativistic effects are especially important. A special
feature of an accretion disk around the rapidly rotating black hole is that 
the viscously dissipated power becomes comparable to
$\dot{M}c^2$, $\dot{M}$ being the accretion rate. As a result, 
the inertial mass associated with heat stored in the advective 
disk is approximately equal to the rest mass of the accreting gas.
Lasota's equations, being fully relativistic in all other respects,
neglect the inertia of internal heat, 
and Peitz \& Appl (1997) and Jaroszy{\'n}ski \& Kurpiewski (1997)
have partly included it in their calculations. In this note we 
show that accounting for the heat inertia is necessary to make the disk
model consistent with the global energy conservation law, and
derive the disk structure equations including this effect. 
In particular, an additional term 
appears in the radial Euler equation. We dub it "heat deceleration" as
it describes the back reaction of energy release on the radial velocity of 
accreting matter.

\section{Notation, assumptions and methods}

Our approach and notation are very close to those in Abramowicz {\it et al.}
(1996), except that we do not neglect the contribution of internal energy and 
pressure to the inertial mass of the flow.
We denote by
$r_g=2GM/c^2$ the gravitational radius of the black hole with the
mass $M$, and by $a = J/Mc$ the Kerr parameter connected to the black
hole angular momentum $J$. 
We use the
Boyer-Lindquist coordinates $x^i = (t, r, \theta, \varphi)$, and the
signature $(- + + +)$. The metric tensor $g_{ij}$ is
given {\it e.g.} by Misner, Thorne, \& Wheeler (1973). 
The four-velocity of the accreting gas has components
$u^i =(u^t, u^r, u^{\theta}, u^{\varphi})$ in the Boyer-Lindquist
coordinates. The disk is assumed to lie at the equatorial plane of the 
Kerr geometry. In the slim approximation, 
$u^\theta$ is neglected and all the metric and connection coefficients are
evaluated at the equatorial plane.

The gas Lorentz factor measured in the frame of local observers with zero
angular momentum is connected to $u^t$ by $\gamma = u^t(-g^{tt})^{-1/2}$. 
The angular velocity of the gas rotation is defined as
$\Omega = u^\varphi/u^t$.
It is not equal in general to the Keplerian angular velocity, 
$$
\Omega_K^\pm=\pm\frac{c}{r(2r/r_g)^{1/2}\pm a}. 
$$
(Plus and minus signs in this formula correspond to co-rotating and
counter-rotating orbits). The $r$-component of four-acceleration, $a_r$,
is given by (Abramowicz {\it et al}., 1996)
$$
a_r=\frac{1}{2}\frac{d}{dr}\left(u_ru^r\right) +\frac{1}{2}
\frac{\partial g_{\varphi\varphi}}{\partial r}g^{tt}\gamma^2
\left(\Omega-\Omega_K^+\right)\left(\Omega-\Omega_K^-\right).
$$

The dynamical equations are derived from the conservation
laws $\nabla_i T^i_{~k}=0$, where $\nabla_i$ is the covariant derivative
in the Kerr metric, and $T^i_{~k}$ is the stress-energy tensor of the viscous 
gas flow (Misner et al. 1973), 
$$
 T^i_{~k}=(\rho+p)u^iu_k+p\delta^i_{~k}-2c\eta\sigma^i_{~k}+
       \frac{1}{c}(u^iq_k+u_kq^i),
$$
where $\rho$ and $p$ are the total energy density and pressure in the 
comoving frame, $\sigma^i_{~k}$ is the shear tensor, and $q^i$ is the energy
flux vector assumed to be directed vertically in the disk.
The dynamic viscosity $\eta$ is related to the kinematic viscosity $\nu$ 
by $\eta=\nu(\rho+p)/c^2$. (Note that this differs from the previously used 
relation that includes only the rest mass of the gas).

The final equations are written for  
the vertically integrated thermodynamic quantities:
surface rest mass density $\Sigma$, surface energy density $U$ 
(that includes both the rest mass energy $\Sigma c^2$, and internal energy 
$\Pi$), and the vertically integrated pressure $P$. 
$F^+$ denotes the surface rate of viscous heating, and 
$F^-$ denotes the radiation flux radiated from both faces of the disk.
All the thermodynamic
quantities, and both fluxes $F^-, F^+$ are measured in the comoving frame. 
The dimensionless specific enthalpy is defined as,
$$
\mu=\frac{U+P}{\Sigma c^2} = 1+\frac{\Pi+P}{\Sigma c^2}.
$$
Neglecting the internal heat contribution to the inertia of the flow
is the same as assuming that $\mu = 1$. In this paper we do not assume that, 
but keep in all equations $\mu>1$. Consistently, we keep the viscous 
term in the radial Euler equation (see Section 4).

We skip all computational details, because our derivation goes along
the same standard lines explained {\it e.g.} by Page \& Thorne (1974),
and used recently by Lasota (1994) and all the other above quoted
subsequent authors. In the next Sections we give only the final 
results.

\section{Conservation of energy and angular momentum}

The conservation of energy and angular momentum are expressed by
equations
\begin{equation}
  \frac{d}{dr}\left[\mu\left(\frac{\dot{M}u_t}{2\pi}+2\nu\Sigma r 
  \sigma_{~t}^r\right)\right]=\frac{F^-}{c^2}\;ru_t,
\end{equation}
\begin{equation}
  \frac{d}{dr}\left[\mu\left(\frac{\dot{M}u_\varphi}{2\pi}
  +2\nu\Sigma r \sigma_{~\varphi}^r\right)\right]=\frac{F^-}{c^2}\;ru_\varphi,
\end{equation}
Here ${\dot M}$ is the accretion rate related to $u^r$ and $\Sigma$ by
the barion conservation law,
\begin{equation}
  2\pi r c u^r \Sigma =-\dot{M},
\end{equation}
and $\sigma_{~t}^r$, $\sigma_{~\varphi}^r$ are the shear tensor components, 
$$
  \sigma_{~\varphi}^r= \frac{1}{2}\;g^{rr}g_{\varphi\varphi}\sqrt{-g^{tt}}
  \gamma^3\frac{d\Omega}{dr}, \qquad
  \sigma_{~t}^r= -\Omega\;\sigma_{~\varphi}^r.
$$
These equations differ from those with neglected inertia of heat 
(see Abramowicz {\it et al.} 1996) just by the presence of the factor 
$\mu > 1$. 
They also differs from the equations in Peitz \& Appl (1997) and
Jaroszy{\'n}ski \& Kurpiewski (1997)
who take into account the contribution of heat to the specific orbital energy 
and angular momentum of the flow but neglect the similar term in the 
relation between the dynamic and kinematic viscosity. 

In the standard thin disk
$\mu\approx 1$ with high accuracy, and the rotation is very close to
Keplerian. Then the equations (1,2) become the same as those derived by
Page \& Thorne (1974). For the advective disk, however, the factor $\mu$
should not be neglected. This can be easily seen from the global energy
conservation law. Integrating equation (1) from the transonic inner edge
of the disk, $r_{in}$, 
to infinity, and neglecting the viscous stress at $r_{in}$, we get
\begin{equation}
    L=-\frac{2\pi}{c}\int\limits_{r_{in}}^\infty u_t F^- rdr=\dot{M}c^2
    \left(1+\mu_{in}\frac{u_t^{in}}{c}\right).
\end{equation}
It follows that the radiative efficiency of the disk equals 
\begin{equation}
\epsilon = 1 - \mu_{in}\left(1-\frac{b_{in}}{c^2}\right),
\end{equation}
where $b_{in}=c^2+u_t^{in}c$ is the specific binding energy at the
inner edge. In the advective
disks, most of the dissipated binding energy is stored as internal heat and
the radiative losses are negligible, $\epsilon\rightarrow 0$.
>From this we conclude that
$\mu_{in}\rightarrow (1-b_{in}/c^2)^{-1}$.
In the particular case of the extremely rotating black hole,
assuming {\it e.g.} that 
$b_{in}/c^2$ is close to the corresponding standard value 
$(1-1/\sqrt{3})\approx 0.42$, 
one has $\mu_{in}=\sqrt{3}$.  Note that the usual assumption
$\mu=1$ follows that $b_{in}=0$ for any advective disk.
In fact, the position and binding energy of the inner edge can be found
only by the integration of the disk structure equations, and 
it is the difference $(\mu-1)$ that adjusts to keep $\epsilon\approx 0$.
In this respect the heat inertia is essential in the case of non-rotating
black hole as well as extremely rotating.

It would be instructive to compare the angular momentum equation (2) 
with its commonly used standard version (Novikov \& Thorne 1973; 
Page \& Thorne 1974). While $\mu = 1$ can be safely assumed in the
standard disk, the radiative losses of angular momentum represented by
the right hand side of equation (2) become important 
when the accreting black hole is rapidly rotating ({\it e.g.} Lamb 1996).
Contrary to this situation, in the advection dominated disk the radiative
losses are always small, but the deviation of $\mu$ from
unity becomes significant. In a sense, the angular momentum which would
be radiated away by the standard disk remains now in place, being
carried by the increased inertia of internal energy.

\section{Radial motion and viscous heating}

To clarify the role of heat inertia in the equation of radial motion we
derive this equation in two ways. Firstly, we perform vertical
integration of the $r$-component of the equation $\nabla_i T^i_{~k}=0$,
to get
\begin{equation}
  u^ru_r\left[\frac{d}{dr}(\Pi+P)-
	      \xi\frac{\Pi+P}{\Sigma}\frac{d\Sigma}{dr}\right]+
  a_r(U+P)+\frac{dP}{dr}+u_r\frac{F^-}{c}=0,
\end{equation}
where $\xi\approx 1$ is a numerical factor accounting for non-homogeneity
of the disk in vertical direction. In this equation we have taken into
account that the divergency of the vertical energy flux $q^i$
becomes equal to $F^-$ after the vertical integration.

Secondly, we project the equation $\nabla_iT^i_{~k}=0$ onto the hypersurface 
orthogonal to $u^i$ to get the relativistic Eurler equation 
({\it c.f.} Lightman et al. 1979, problem 5.31).
Then the flux term vanishes, but the viscous term arises in the 
radial equation, and we get after the vertical integration
\begin{equation}
   a_r(U+P)+\frac{dP}{dr}(1+u^ru_r)+\frac{F^+}{c} u_r=0,
\end{equation}
where 
\begin{equation}
  F^+=2\nu\Sigma \mu\; \sigma^2 c^2, \qquad
  \sigma^2=\frac{1}{2} g^{rr}g_{\varphi\varphi}\left(-g^{tt}\right)
  \gamma^4\left(\frac{d\Omega}{dr}\right)^2.  
\end{equation}

Keeping $F^\pm$ terms in the equations (6,7) allows to get the first
law of thermodynamics for a vertically integrated accretion disk as a
consequence of these two equations,
\begin{equation}
   F^+-F^-=cu^r\left(\frac{d\Pi}{dr}
   -\xi \frac{\Pi+P}{\Sigma}\frac{d\Sigma}{dr}\right).
\end{equation}

Finally, substituting $a_r$ to (7), we write down the 
equation of radial motion with $u^ru_r\ll 1$,
\begin{equation}
   \frac{1}{2}\;\frac{d}{dr}\left(u_ru^r\right)=
   -\frac{1}{2}\;\frac{\partial g_{\varphi\varphi}}{\partial r}
   g^{tt}\gamma^2
   \left(\Omega-\Omega_K^+\right)\left(\Omega-\Omega_K^-\right)
  -\frac{1}{c^2\Sigma \mu}\frac{dP}{dr}-\frac{F^+u_r}{c^3\Sigma \mu}.
\end{equation}
This equation differs from the radial equation in Abramowicz
{\it et al.} (1996) by the presence of the factor $\mu$, and by the
additional term proportional to $F^+$. Without this term, equation (10) 
would have a simple meaning: radial acceleration is a combined result of 
the deviation from Keplerian rotation and the radial pressure gradient.  
The viscous term $\propto F^+$ represents the heat deceleration effect
which can be interpreted in the following
way. The mass of the accreting gas measured in its local rest frame 
increases due to the stored heat. Thus, from the local
observer point of view, the dissipation of orbital binding energy is an
external source of mass-energy, and the matter inflow proceeds like 
motion of a body with changing mass. In this case, in addition to acting 
forces there is a contribution to acceleration due to the change of mass.
The mass of the accreting gas increases with rate proportional to $F^+$
and this tends to decelerate the matter inflow. The "deceleration by 
heating" is a purely relativistic effect as it accounts for the 
mass-energy relation. It is proportional to the ratio of the released 
power to $\dot{M}c^2$ and can be essential near a 
rapidly rotating black hole. Then the additional term must influence 
the structure of the advective disk in the innermost region. There it becomes 
of the same order as the left hand side of the equation (10).

Note that the deceleration term is independent of the radiative losses,
$F^-$, and remains the same even when $F^-\sim F^+$. In this case the bulk of
the dissipated energy is radiated away. However, this energy first appears as 
heat decelerating the radial inflow. Only then it is radiated away.
The net momentum taken away by the radiation flux vanishes in the rest
frame of the accreting gas, and does not contribute to the velocity
change.

\section*{Acknowledgements}

This work was supported in part by the Danish National Research Foundation 
through its establishment of the Theoretical Astrophysics Center,
by the Danish Natural Science Research Council through grant 11-9640-1,
and by the Nordita's Nordic Project {\it Non-linear phenomena in accretion 
disks around black holes}.
AMB thanks the Theoretical Astrophysics Center and the Department of 
Astronomy \& Astrophysics at Chalmers University of Technology 
for hospitality, and acknowledges partial support by RFFI grant 97-02-16975.

\newpage

\section*{References}
\parindent=0pt

Abramowicz, M.A., Chen, X.-M., Granath, M., \& Lasota, J.-P. 1996, 
ApJ, 471, 762

Abramowicz, M.A., Czerny, B., Lasota, J.-P., \& Szuszkiewicz, E. 1988, 
ApJ, 332, 646

Abramowicz, M.A., Lanza, A., \& Percival, M.J. 1997, ApJ, 479, 179

Igumenshchev, I.V., Novikov, I.D., \& Abramowicz, M.A. 1997, in preparation

Jaroszy{\'n}ski, M., \& Kurpiewski, A. 1997, A\&A, in press

Lamb, F. 1996, in Basic Physics of Accretion Disks, ed. S. Kato,
S. Inagaki, J. Fukue, \& S. Mineschige \\
\hspace*{0.5cm}(New York: Gordon \& Breach), in press

Lasota, J.-P. 1994, in Theory of Accretion Disks-2, ed. W.J. Duschl, 
J. Frank, F. Meyer,\\ 
\hspace*{0.5cm} E. Meyer-Hofmeister, \& W.M. Tscharnuter (Dordrecht: 
Kluwer), 341

Lightman, A.P., Press, W.H., Price, R.H., \& Teukolsky, S.A. 1979,
Problem Book in Relativity and\\ 
\hspace*{0.5cm}Gravitation (Princeton University Press)

Misner, C.W., Thorne, K.S., \& Wheeler, J.A. 1979, Gravitation
(San Francisco: Freeman)

Narayan, R. 1996, astro-ph/9611113

Novikov, I.D., Thorne, K.S. 1973, in Black Holes, 
ed. C. de Witt \& B.S. de Witt \\
\hspace*{0.5cm}(New York: Gordon \& Breach), 343

Paczy{\'n}ski, B., Bisnovatyi-Kogan, G.S. 1981, Acta Astron., 31, 3

Page, D.N., Thorne, K.S. 1974, ApJ, 191, 499

Peitz, J., Appl, S. 1997, MNRAS, 286, 681 

Shakura, N.I. 1972, Soviet Astron., 16, 756

Shakura, N.I., Sunyaev, R.A. 1973, A\&A, 24, 337

Szuszkiewicz, E., Malkan, M.A., \& Abramowicz, M.A. 1996, ApJ, 458, 474

\end{document}